\documentclass[
 reprint,
 longbibliography,
 amsmath,amssymb,
 aps,
prl,
floatfix,superscriptaddress
]{revtex4-1}
\usepackage[colorlinks=true,linkcolor=blue,urlcolor=blue,citecolor=blue]{hyperref}    
\usepackage{cleveref}
\usepackage{enumerate}
\usepackage[caption=false,singlelinecheck=off]{subfig}
\usepackage{slashed}
\usepackage{natbib}
\usepackage{braket}
\usepackage{bbold}
\usepackage{booktabs}
\usepackage{blindtext}
\usepackage{multibib}
\usepackage{bbm}
\usepackage{epsfig}
\usepackage{graphicx}
\usepackage{dcolumn}
\usepackage{bm}

\begin{document}
\title{Topological Classification of Shot Noise on Fractional Quantum Hall edges}
\author{Christian Sp\r{a}nsl\"{a}tt}
\affiliation{Institut f\"{u}r Nanotechnologie, Karlsruhe Institute of Technology, 76021 Karlsruhe, Germany}
\affiliation{Institut f\"{u}r Theorie der Kondensierte Materie, Karlsruhe Institute of Technology, 76128 Karlsruhe, Germany}
\author{Jinhong Park}
\affiliation{Department of Condensed Matter Physics, Weizmann Institute of Science, Rehovot 76100, Israel}
\author{Yuval Gefen}
\affiliation{Institut f\"{u}r Nanotechnologie, Karlsruhe Institute of Technology, 76021 Karlsruhe, Germany}
\affiliation{Department of Condensed Matter Physics, Weizmann Institute of Science, Rehovot 76100, Israel}
\author{Alexander D. Mirlin}
\affiliation{Institut f\"{u}r Nanotechnologie, Karlsruhe Institute of Technology, 76021 Karlsruhe, Germany}
\affiliation{Institut f\"{u}r Theorie der Kondensierte Materie, Karlsruhe Institute of Technology, 76128 Karlsruhe, Germany}
\affiliation{Petersburg Nuclear Physics Institute, 188300 St. Petersburg, Russia}
\affiliation{L.\,D.~Landau Institute for Theoretical Physics RAS, 119334 Moscow, Russia}

\date{\today}
\begin{abstract}
	Electrical and thermal transport on a fractional quantum Hall edge are determined by topological quantities inherited from the corresponding bulk state. While electrical transport is the standard method for studying edges, thermal  transport appears more challenging. Here, we show that the shot noise generated on the edge provides a fully electrical method to probe the edge structure. In the incoherent regime, the noise falls into three topologically distinct universality classes: charge transport is always ballistic while thermal transport is either ballistic, diffusive, or ``antiballistic''. Correspondingly, the noise either vanishes, decays algebraically or is constant up to exponentially small corrections in the edge length.
	\end{abstract}
\maketitle

Electrical and thermal transport on fractional quantum Hall (FQH) edges are quantized, reflecting the bulk topological order~\cite{Wen1990a,Wen1990b,Chang2003}. In particular, the electrical Hall and two-terminal conductances, $G_H$ and $G$,  
\begin{equation}
G_H = \nu e^2 / h, \qquad G = |G_H|,
\label{eq:G}
\end{equation}
are determined by the bulk filling factor $\nu\in \mathbb{Q}$. Furthermore, the thermal Hall and two-terminal conductances, $G^Q_H$ and $G^Q$, 
\begin{equation}
G^Q_H = \nu_Q\kappa T, \qquad G^Q = |G^Q_H|,
\label{eq:GQ}
\end{equation}
manifest  the difference $\nu_Q\equiv n_d-n_u\in \mathbb{Z}$ between the number of ``downstream'' $n_d$ and ``upstream'' $n_u$ 
(with respect to the chirality direction set by the magnetic field)
edge channels~\cite{Kane1997,Capelli2002}; for Abelian states  $\nu_Q \in \mathbb{Z}$. Here $T$ is the temperature, $\kappa = \pi^2 k_B^2/(3h)$,  and $k_B$ is Boltzmann's constant. It should be emphasized that, for edges with counterpropagating channels, the quantizations~\eqref{eq:G} and~\eqref{eq:GQ} hold under the condition (and is a hallmark) of the incoherent regime \cite{Protopopov2017,Nosiglia2018}. This regime is generic for nearly all FQHE experiments on such edges; reaching the coherent regime requires very low temperatures and/or very small distances between electrodes and/or special control over intermode tunneling \cite{Cohen2019}. 

Since the discovery of the FQH effect~\cite{Stormer1982,Laughlin1983}, electrical transport has been the standard method for probing the edge structure. On the other hand, experiments on thermal transport have only recently determined the number of active channels for a variety of edges~\cite{Jezouin2013,Banerjee2017,Banerjee2018}. These experiments are technically very challenging, and so far only capable of determining $|\nu_Q|$. The heat conductance was measured recently also for a quantum spin liquid~\cite{Kasahara2018}, which belongs to a class of topologically ordered states distinct from the FQH states.

In this Letter,  we classify FQH edges into  topological classes according to the scaling of the shot noise $S$ with the edge length $L$. We thus show that a measurement of the noise is an efficient diagnostic tool for determining important features of the edge structure. We emphasize that this method of probing the edge structure is purely electrical and thus \textit{complementary} to the (very difficult) measurements of the heat conductance. 

Noise measurements are already a ubiquitous tool in mesoscopic physics~\cite{deJong1996,Blanter2000}, in particular for FQH experiments, where they are used to probe fractionally charged quasiparticles in quantum point contact (QPC) geometries~\cite{Saminadayar1997,DePicciotto1997,Dolev2008}. Shot noise has also been suggested as a probe of quasiparticle statistics~\cite{Feldman2007}.

We focus on the incoherent transport regime (which is generic as emphasized above)  $L/l_{\rm eq}\gg 1$, where  $l_{\rm eq}$ is a characteristic equilibration length.
Beyond $l_{\rm eq}$, individual channels cannot be distinguished as they fully equilibrate and form hydrodynamic modes. By the same token, any effects from edge reconstruction~\cite{Meir1994,Wang2013} are negligible. What remains are the transport coefficients $\nu$ and $\nu_Q$, which can be observed either through conductance measurements or also, as we emphasize in this Letter, in the shot noise. The incoherent regime has been discussed in early works~\cite{Kane1995,Kane1997} and more recently studied in the context of line junctions~\cite{Sen2008,Rosenow2010} and of the $\nu=2/3$ edge~\cite{Protopopov2017,Nosiglia2018,Aharon2019,Park2019}.

\begin{figure}[t]
\includegraphics[width=1.0\columnwidth] {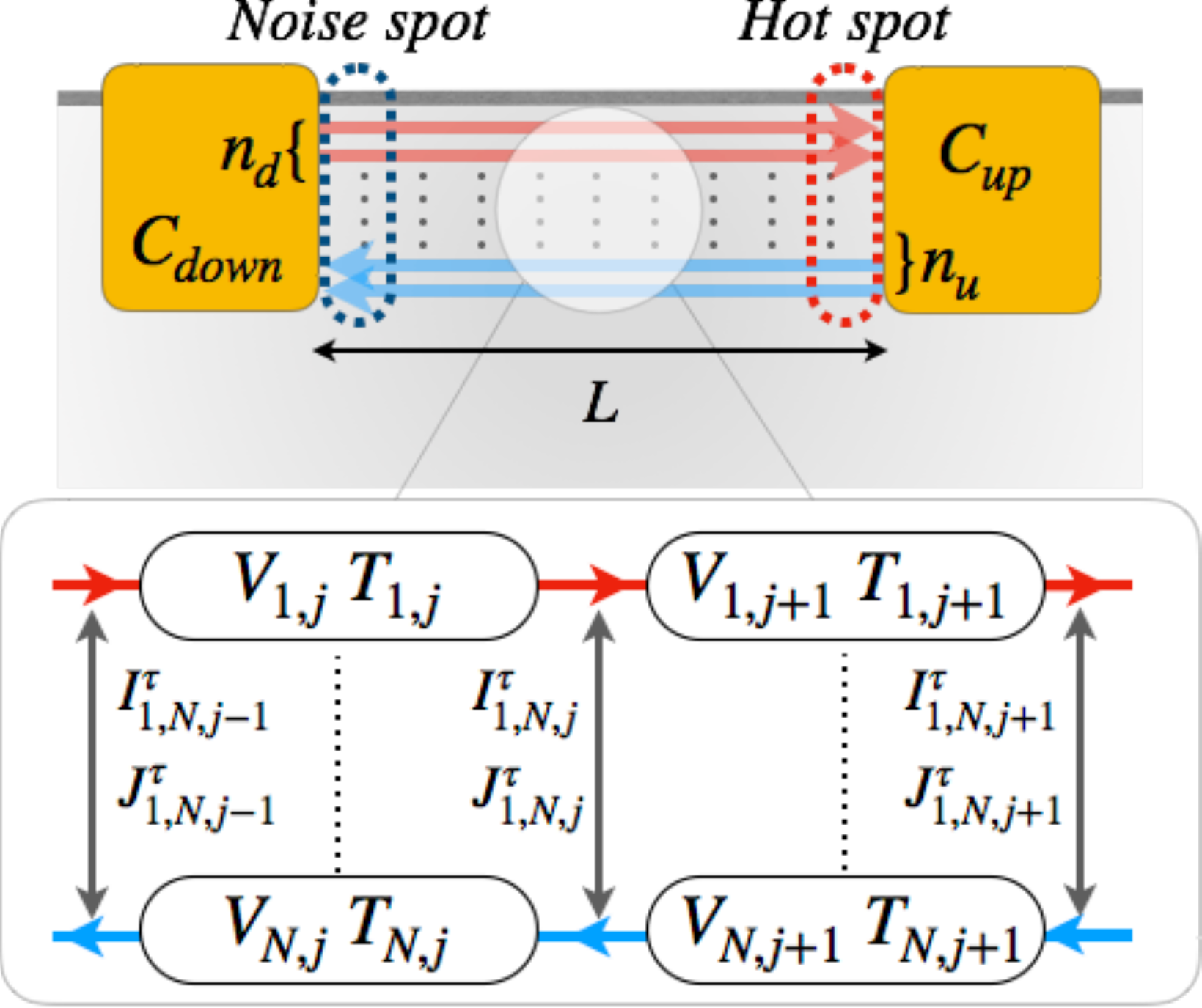}\caption{\label{fig:Model} Model of equilibration on the FQH edge: a set of $n_d$ downstream modes (red) and $n_u$ upstream modes (blue) emanate from contacts $C_{\rm down}$ and $C_{\rm up}$ (separated by distance $L$), respectively, and equilibrate due to intermode disorder-mediated tunneling of charge and energy.  Local voltages $V_{n,j}$ and temperatures $T_{n,j}$ are determined by local virtual reservoirs (labeled by $j$), which drive charge $I^{\tau}_{n,m,j}$ and energy $J^{\tau}_{n,m,j}$  tunneling currents between channels $n \leftrightarrow m$. (The reservoirs do not absorb charge or energy.) Our proposed measurement scheme consists of voltage biasing one contact and measuring noise. Heating due to voltage equilibration then occurs only in the vicinity of $C_{\rm up}$, the \textit{hot spot}, and heat spreads along the edge. The type of heat conduction is fixed by the topological quantity $\nu_Q \equiv n_d-n_u$. The shot noise measured in any of the contacts is effectively Johnson-Nyquist noise generated  in a region close to $C_{\rm down}$: \textit{the noise spot}. This noise is independent of which direction the bias is put (left-to-right or right-to-left).}
\end{figure}

We consider a general Abelian FQH edge (see Fig.~\ref{fig:Model}) and show that an edge with $\nu_Q>0$ generates vanishing noise $S\simeq 0$ (up to exponential corrections in $L/l_{\rm eq}$), $\nu_Q=0$ generates $S\sim \sqrt{l_{\rm eq}/L}$, and edges with $\nu_Q<0$ generate $S\simeq {\rm const}$. These three types of noise asymptotics reflect respectively the three possible combinations of electrical and thermal transport: \textit{(i)} Both charge and heat flows ballistically downstream. This is for instance the case for the integer quantum Hall states~\cite{Klitzing1980} as well as for particlelike states such as $\nu=1/3$, $\nu=2/5$, etc. \textit{(ii)} Charge flows downstream, but the heat transport is diffusive. This holds whenever $\nu_Q=0$, with $\nu=2/3$ as the most studied example~\cite{MacDonald1990,Johnson1991,Kane1994,Protopopov2017,Nosiglia2018,Park2019}. \textit{(iii)} Charge flows downstream, but heat flows ballistically upstream; we term such heat transport ``antiballistic''. This is the case for hole-conjugate states such as, e.g., $\nu = 3/5$, $\nu = 4/7$, etc, which are characterized by $n_d < n_u$~\cite{Kane1995b}. 

The mechanism for such strikingly different noise characteristics follows from a spatial separation of where heat and noise is generated~\cite{Park2019}. This separation is a consequence of the chiral nature of the edge which, in turn, is a manifestation of the $U(1)$ gauge anomaly in $2+1D$ Abelian Chern-Simons theory~\cite{Callan1985,Stone1991,Wen1995}.

More concretely, full equilibration on the edge yields a two-terminal charge conductance $G = G_d + G_u \simeq\nu e^2/h$, where
$G_d \simeq\nu e^2/h$ and $G_u \simeq 0$ describe the current on the considered portion of the edge in response to the voltages on the electrodes $C_{\rm down}$ and $C_{\rm up}$, respectively~\cite{SuppMat}. Regardless of which of the two electrodes is a source (respectively, a drain), the local voltage along the edge drops only within a distance $\sim l_{\rm eq}$ \textit{in the vicinity of $C_{\rm up}$} (see Fig.~\ref{fig:Model}) and is therefore fixed by the net chirality (i.e, by the downstream direction). The region of the voltage drop is associated with Joule heating, and is therefore referred to as the \textit{hot spot}~\cite{Tamaki2005,Park2019}. The nature and direction of heat transport away from the hot spot depend crucially on $\nu_Q$ which is fixed by the topological order of the bulk~\cite{Kane1997,Capelli2002}.

Because of the heating along the edge, thermally activated tunneling between the edge channels can excite particle-hole pairs. Such pairs exist only within $\sim l_{\rm eq}$ from where they are generated; beyond that distance they recombine due to equilibration processes. If the respective constituents of a pair would reach different contacts, fluctuations in the charge current---shot noise---will be observed in the contacts. Crucially, only pairs created within a distance $\sim l_{\rm eq}$ of $C_{\rm down}$ have a non-negligible probability to contribute to this type of partition noise. Particles or holes further away from this point will experience equilibration due to repeated tunneling and, as a result, will 
both flow in the downstream direction, finally reaching $C_{\rm up}$. Hence, the region $\sim l_{\rm eq}$ close to $C_{\rm down}$ is referred to as the \textit{noise spot}. Interestingly, this noise generating mechanism carries some resemblance to the photon pair splitting picture of Hawking radiation~\cite{GravitationBook}.

\textit{Model.--} We consider a segment of a general Abelian FQH edge with $N$ edge channels between two contacts separated by a distance $L$ (see Fig.~\ref{fig:Model}). For each channel $n$, corresponding to a filling factor discontinuity $\nu_n$, we introduce $M$ virtual reservoirs (which absorb neither charge nor energy) modeling local equilibration~\cite{Nosiglia2018,Park2019,Engquist1981}. For simplicity, we neglect any temporary charge or heat accumulation in the reservoirs (in any case, such an accumulation would not affect the zero frequency noise). Hence, at any time $t$, charge and energy currents are locally conserved
\begin{subequations}
\label{eq:ConservationReservoirs}
\begin{align}
& I_{n,j,{\rm out}}(t) = I_{n,j,{\rm in}}(t) \equiv I_{n,j}(t), \\
& J_{n,j,{\rm out}}(t) = J_{n,j,{\rm in}}(t) \equiv J_{n,j}(t).
\end{align}
\end{subequations}
Here, $I_{n,j,{\rm out(in)}}(t)$ is the outgoing (incoming) charge current of channel $n$ into its reservoir at location $j$, and $J_{n,j,{\rm out(in)}}(t)$ are corresponding energy currents. Since we are only interested in steady state conditions and zero frequency noise, all quantities are hereafter assumed to be time averaged and we therefore drop any explicit $t$ dependence. The local voltages $V_{n,j}\equiv (h/e^2 \chi_n \nu_n) I_{n,j}$, where $\chi_n=\pm 1$ is the channel chirality, and temperatures $T_{n,j}$ drive charge and energy tunneling currents $I^\tau_{n,m,j}$ and $J^\tau_{n,m,j}$ from channel $n$ to $m$ at position $j$,
\begin{subequations}
\label{eq:ConservationTunneling}
\begin{align}
&I_{n,j+1} = I_{n,j} - \sum_{m=1}^N I^\tau_{n,m,j},\\
&J_{n,j+1} = J_{n,j} - \sum_{m=1}^N J^\tau_{n,m,j}.
\end{align}
\end{subequations}
To lowest order in the tunneling couplings $g_{n,m}\equiv g_{m,n}\ll1$~\cite{gDepNote}, these currents read
\begin{subequations}
\label{eq:TunnelingCurrents}
\begin{align}
I^\tau_{n,m,j} = - I^\tau_{m,n,j} &=g_{n,m}\frac{e^2}{h}\left(V_{n,j}-V_{m,j}\right),\\
J^\tau_{n,m,j} = - J^\tau_{m,n,j} &=g_{n,m}\frac{e^2}{2h}\left(V_{n,j}^2-V^2_{m,j}\right) \notag\\ 
& + \gamma_{n,m} g_{n,m}\frac{\kappa}{2} \left(T^2_{n,j}-T^2_{m,j}\right),
\end{align}
\end{subequations}
where $\gamma_{n,m} \equiv \gamma_{m,n}$ parametrize the deviation (of order unity) from Wiedemann-Franz law for the interchannel tunneling current. In general, $\gamma_{n,m}$ 
depends on $\nu_n$, $\nu_m$, and the intermode interactions. As an example, in the case of noninteracting channels with $\nu_n = 1$ and $\nu_m = 1/(2p+1)$ with integer $p$ one finds $\gamma_{n,m}$ = $3/(2\nu_m+1)$~\cite{Nosiglia2018}. 

By combining Eqs.~\eqref{eq:ConservationTunneling} and~\eqref{eq:TunnelingCurrents}, we derive continuum equations for the voltage, current, and temperature profiles along the edge. We denote by $a$ the distance between successive reservoirs and define $x=ja$. Taking the continuum limit $g_{n,m}\rightarrow 0$, $a\rightarrow 0$, $M\rightarrow \infty$, with the coordinate $x$ and the equilibration length $l_{n,m}\equiv a/g_{n,m}$ kept finite, we obtain the following equation for the local voltages:
\begin{equation}
\label{eq:VoltageEquation}
\partial_{x}\vec{V}(x) = \mathcal{M}_V \vec{V}(x),
\end{equation}
where $\vec{V}(x)=(V_1,\hdots,V_N)^T(x)$ (superscript $T$ denotes transposition) and 
\begin{equation}
\label{eq:VMatrix}
\mathcal{M}_V = \begin{pmatrix}
-\frac{\sum_{n\neq1}l^{-1}_{1,n}}{\chi_1 \nu_1} & \frac{l^{-1}_{1,2}}{\chi_1\nu_1} & \hdots & \frac{l^{-1}_{1,N}}{\chi_1\nu_1}\\
\frac{l^{-1}_{1,2}}{\chi_2 \nu_2} & -\frac{\sum_{n\neq 2} l^{-1}_{2,n}}{\chi_2 \nu_2} &\hdots & \frac{l^{-1}_{2,N}}{ \chi_2 \nu_2}\\
\vdots & \vdots & \ddots & \vdots  \\
\frac{ l^{-1}_{1,N}}{  \chi_N\nu_N} & \frac{ l^{-1}_{2,N}}{ \chi_N \nu_N} &\hdots & -\frac{\sum_{n\neq N} l^{-1}_{N,n}}{  \chi_N\nu_N}
\end{pmatrix}.
\end{equation}
The local electric currents $\vec{I}(x)$ obey a similar equation,
\begin{equation}
\label{eq:CurrentEquation}
\partial_{x}\vec{I}(x) = \mathcal{M}_I \vec{I}(x), \qquad \mathcal{M}_I =   \mathcal{D} \mathcal{M}_V  \mathcal{D}^{-1},
\end{equation}
 with  $\mathcal{D} = \text{diag}(\chi_1 \nu_1, \hdots,\chi_N \nu_N )$.

For the local temperatures, we obtain
\begin{equation}
\label{eq:TemperatureEquation}
\partial_x\vec{T^2}(x) = \mathcal{M}_T \vec{T^2}(x) + \Delta\vec{V}(x),
\end{equation}
with $\vec{T^2}(x)=(T_1^2,\hdots,T_N^2)^T(x)$ and 
\begin{equation}
\mathcal{M}_T = \begin{pmatrix}
-\frac{\sum_{n\neq1}\tilde{l}^{-1}_{1,n}}{\chi_1} & \frac{\tilde{l}^{-1}_{1,2}}{\chi_1 } & \hdots & \frac{\tilde{l}^{-1}_{1,N}}{\chi_1}\\
\frac{\tilde{l}^{-1}_{1,2}}{\chi_2} & -\frac{\sum_{n\neq 2} \tilde{l}^{-1}_{2,n}}{\chi_2} &\hdots & \frac{\tilde{l}^{-1}_{2,N}}{ \chi_2}\\
\vdots & \vdots & \ddots & \vdots  \\
\frac{ \tilde{l}^{-1}_{1,N}}{  \chi_N} & \frac{ \tilde{l}^{-1}_{2,N}}{ \chi_N} &\hdots & -\frac{\sum_{n\neq N} \tilde{l}^{-1}_{N,n}}{  \chi_N}
\end{pmatrix}.
\end{equation}
Here, we have defined the thermal equilibration lengths $\tilde{l}_{n,m}\equiv l_{n,m}/\gamma_{n,m}$, and
\begin{equation}
\Delta \vec{V}(x) =\frac{e^2}{h\kappa}\sum_{n=1}^N(\frac{(V_1-V_n)^2}{l_{1,n}\chi_1},\hdots,\frac{(V_N-V_n)^2}{l_{N,n}\chi_N})^T(x)
\end{equation}
reflects the Joule heating contribution.

To compute the noise, we consider fluctuations in the charge currents and let $\delta X$ denote the deviation of any quantity $X$ from its time average: $\delta X\equiv X-\overline{X}$.  We decompose the charge tunneling current fluctuations into
\begin{equation}
\delta I^\tau_{n,m,j} = \delta I^{\tau,{\rm int}}_{n,m,j} + \delta I^{\tau,{\rm tr}}_{n,m,j}.
\label{eq:delta-tau}
\end{equation}
Here the intrinsic contributions (superscript ``int'') arise from local Johnson-Nyquist noise; we take them to be independent random variables with zero mean and with variance
\begin{equation}
\label{eq:TempFluct}
\overline{\delta I^{\tau,{\rm int}}_{n,m,j}\delta I^{\tau,{\rm int}}_{n',m',j'}}= \frac{2e^2}{h} g_{n,m}k_B\left(T_{n,j}+T_{m,j'}\right) \delta_{nmj,n'm'j'}.
\end{equation}
We have here assumed that the local voltage difference between any two channels is much smaller than their average temperature: $V_{n,j}-V_{m,j}\ll k_B(T_{n,j}+T_{m,j})/2$. This approximation is excellent (holds up to exponentially small corrections in $L/l_{\rm eq}$) around the noise spot, where the channels equilibrate to the same voltage \cite{SuppMat}. 

The transmitted contributions (superscript ``tr'') in Eq.~\eqref{eq:delta-tau} reflect fluctuations in the voltage difference between the channels 
\begin{equation}
\delta I^{\tau,{\rm tr}}_{n,m,j} = g_{n,m} \frac{e^2}{h}\left(\delta V_{n,j}-\delta V_{m,j}\right),
\end{equation}
that are induced by $\delta I^{\tau,{\rm int}}_{n,m,j}$  according to the transport equation~\eqref{eq:VoltageEquation}. 
In the continuum limit, we find the following equation for the local electric current fluctuations
\begin{equation}
\label{eq:NoiseEquation}
\partial_x\vec{\delta I^{}}(x) = \mathcal{M}_I \vec{\delta I^{}}(x) + \vec{\delta I}^{\tau,{\rm int}}(x),
\end{equation}
where
\begin{equation}
\label{eq:CurrentVariance}
\vec{\delta I}_n^{\tau,{\rm int}}(x) = -\sum_{m>n}^N \delta I^{\tau,{\rm int}}_{n,m}(x)+\sum_{m<n}^N \delta I^{\tau,{\rm int}}_{m,n}(x)
\end{equation}
and $\delta I^{\tau,{\rm int}}_{m,n}(x) \equiv \delta I^{\tau,{\rm int}}_{m,n,j}/a$. With Eqs.~\eqref{eq:VoltageEquation}, \eqref{eq:TemperatureEquation}, and \eqref{eq:NoiseEquation}, we can compute the noise for any edge structure. However, due to the hierarchy of the $N(N-1)/2$ length scales $\sim l_{n,m}$ the noise expressions are algebraically involved. To render the physics more transparent, we therefore employ the following simplification scheme~\cite{SuppMat}: (i) We let all length scales associated with equilibration between channels of the same chirality be $\sim a$. This approximation does not influence the dc noise, since tunneling between copropagating channels does not give rise to any partitioning of the current. (ii) The edge is then transformed into two hydrodynamic modes (consisting of $n_d$ and $n_u$ channels respectively) with effective filling factor discontinuities $\nu_{\pm}$. There remains only a single length scale $l_{\rm eq} \equiv a\nu_+\nu_-/(g_{+,-}(\nu_+-\nu_-))$, where $g_{+,-}$ is the effective tunneling coupling between the modes. Hence,  the treatment of the edge is  reduced to that of two channels. Evaluating the noise, we find~\cite{SuppMat}
\begin{equation}
\label{eq:Noise2ModesMain}
S \simeq \frac{2e^2}{h l_{\rm eq}}\frac{ \nu_-}{ \nu_+} (\nu_+-\nu_-) \int_0^L dx\;\frac{e^{-\frac{2x}{l_{\rm eq}}}k_B\left[T_+(x) +T_-(x) \right]}{(1-e^{-\frac{L}{l_{\rm eq}}}\nu_-/\nu_+)^2},
\end{equation}
where $T_\pm$ are the temperatures of the hydrodynamic modes. It is clear from Eq.~\eqref{eq:Noise2ModesMain} that, while the noise in principle originates from the heating along the full edge, the dominant contribution is given by the region of extension  $\sim l_{\rm eq}$ near $C_{\rm down}$ ($x=0$).  Other contributions  are exponentially suppressed in $x/l_{\rm eq}$.
\begin{figure}[t]
\captionsetup[subfigure]{labelformat=empty}
\subfloat[]{
\includegraphics[width=0.85\columnwidth]{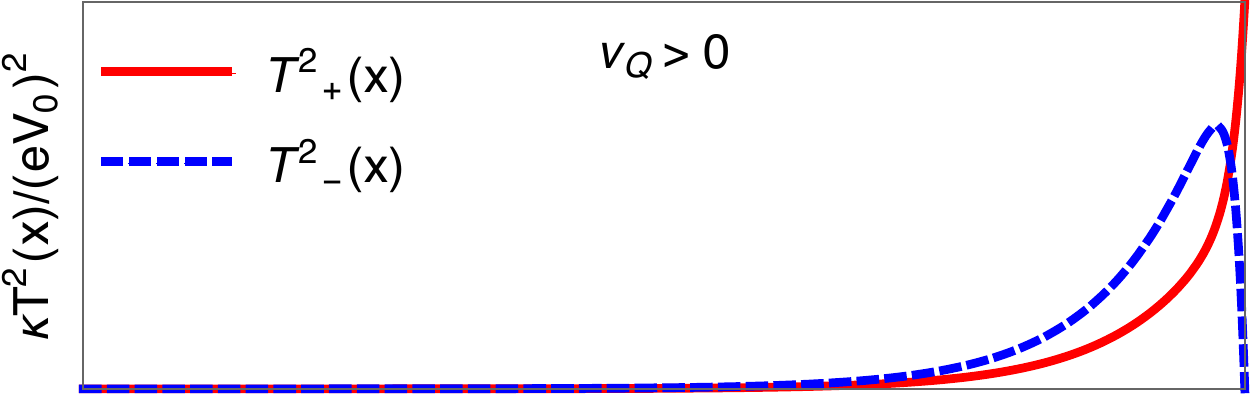}
\label{fig:TB}}
\\[-0.4cm]
\subfloat[]{
\includegraphics[width =0.85\columnwidth]{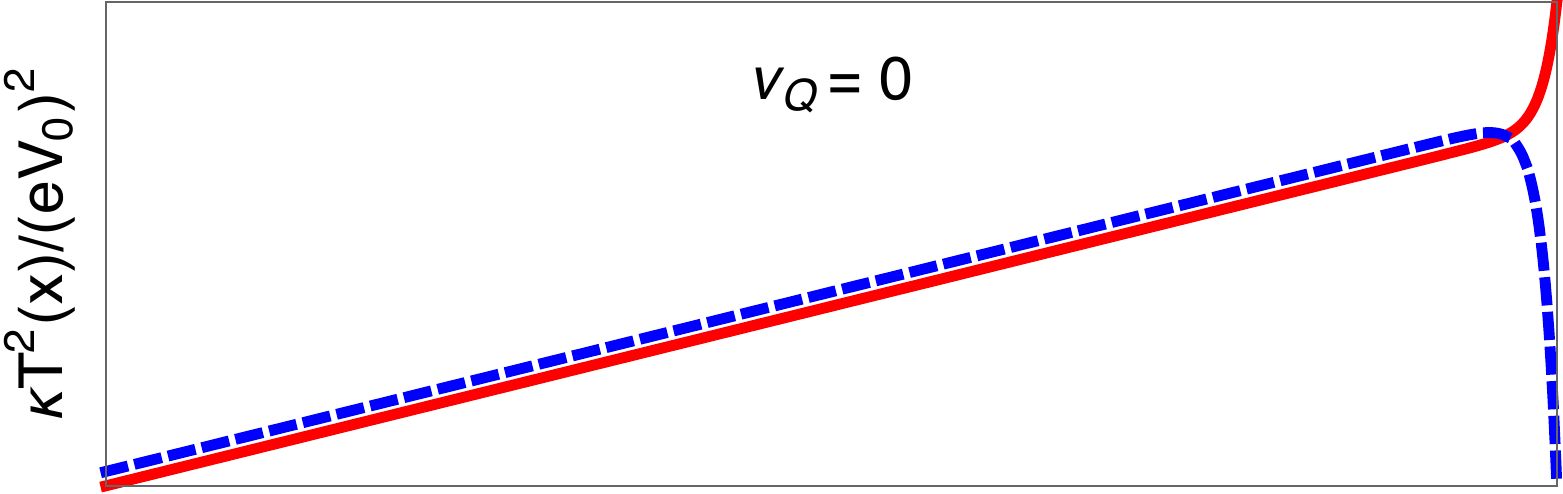}
\label{fig:TD}}
\\[-0.4cm]
\subfloat[]{
\includegraphics[width =0.855\columnwidth]{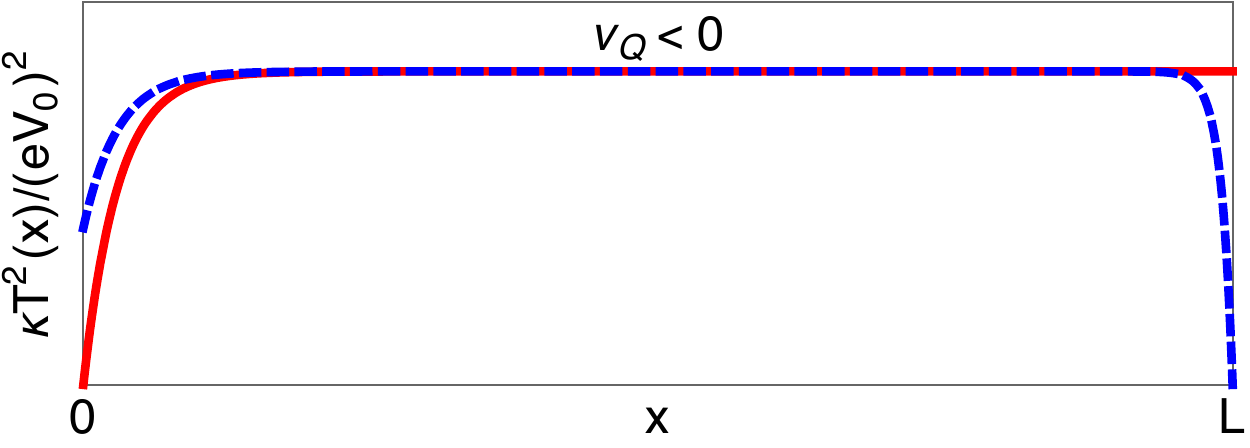}
\label{fig:TAB}}
\caption{Schematics of topologically distinct temperature profiles resulting from equilibration and associated dissipation on FQH edges. $T^2_{\pm}$ are the temperatures of the $n_d$ downstream (+) and $n_u$ upstream modes (-). (Top) $\nu_Q \equiv n_d - n_u>0$, yielding ballistic heat transport downstream. (The case with $n_u\neq 0$ is displayed; for $n_u=0$ the temperature profiles are trivial: $T_{\pm}(x)=0$). (Middle) $\nu_Q=0$, implying heat diffusion. (Bottom) $\nu_Q<0$ with antiballistic (upstream) heat transport.}
\label{fig:Results}
\end{figure}
It follows that the characteristics of the noise depend crucially on the temperature profiles $T_\pm(x)$. Within our approximation, channels of a given chirality are assumed to have the same temperature. We therefore solve Eqs.~\eqref{eq:VoltageEquation} and~\eqref{eq:TemperatureEquation} for two effective hydrodynamic modes, taking the composition of $n_d$ and $n_u$ channels into account~\cite{SuppMat}. The temperature profiles are then determined by the two eigenvalues of the matrix $\exp(x\mathcal{M}_T)$. Their $x$ dependence is crucially determined by $\nu_Q$. We then find, up to boundary effects, that whenever $\nu_Q>0$, $T_\pm(x)$ decays exponentially with $(L-x)/l_{\rm eq}$
(for the special case $n_u=0$, the temperature profiles equal the contact temperature defined to be zero); for $\nu_Q=0$, $T_\pm(x) \sim \sqrt{x/L}$; and for $\nu_Q<0$, $T_\pm(x) \simeq {\rm const}$. We depict these three types of temperature profiles in Fig.~\ref{fig:Results}.

\begin{figure}[t]
\includegraphics[width=1\columnwidth] {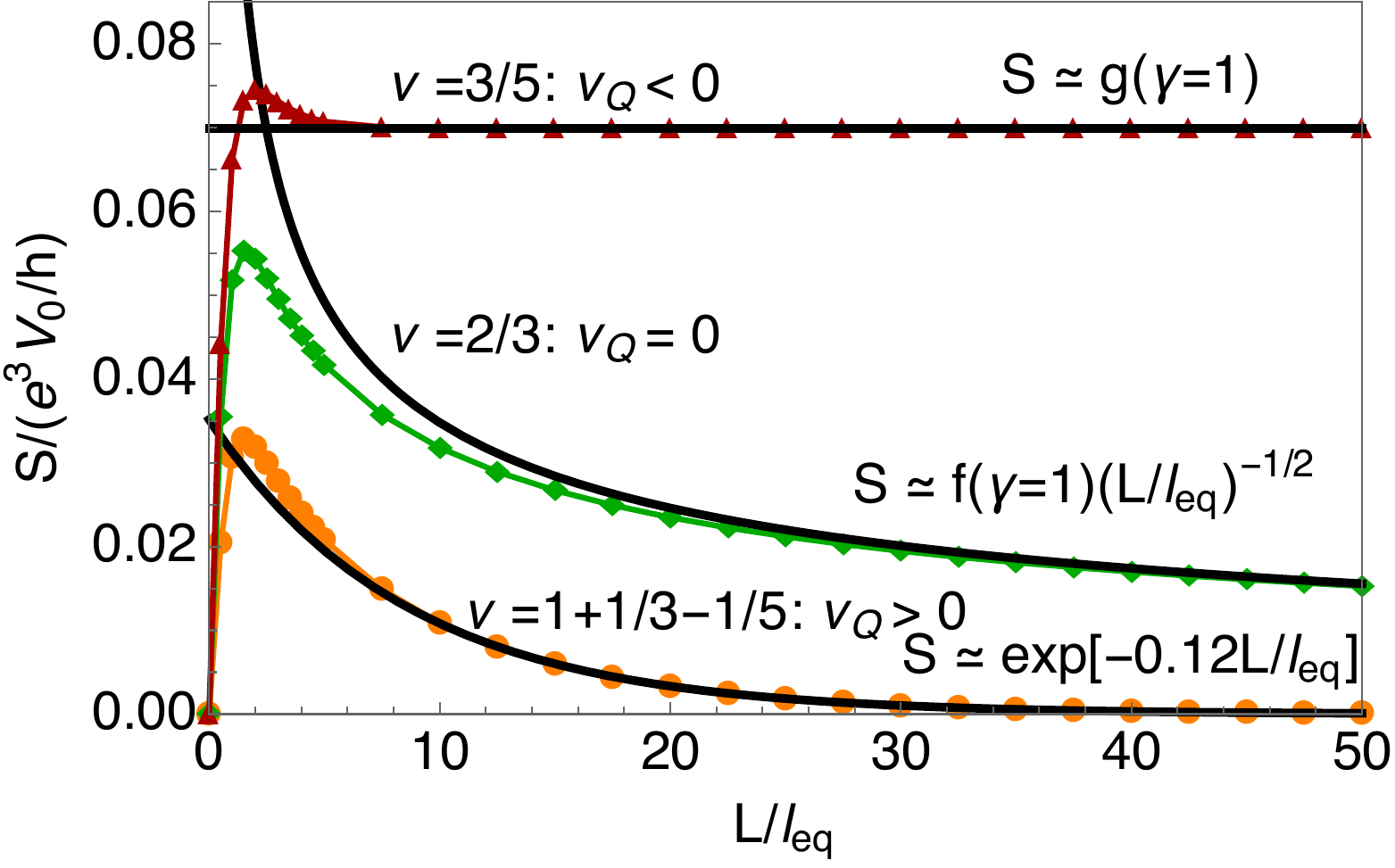}\caption{\label{fig:NoisePlots} Shot noise $S/(e^3V_0/h)$ (where $V_0$ is the bias voltage) on the FQH edge as a function of the edge length $L$ in units of the equilibration length $l_{\rm eq}$ for three distinct edge structures (for simplicity, we have chosen $\gamma=1$ for all three curves). (Yellow circles) Interface between $\nu=4/3$ and $\nu=1/5$ with $\nu_Q >0$  and $S\simeq \exp(-3.35-0.12L/l_{\rm eq})$. (Green diamonds) $\nu=2/3$ with $\nu_Q = 0$ and $S\simeq f(\gamma = 1)\sqrt{l_{\rm eq}/L}$. (Red triangles) $\nu=3/5$  with $\nu_Q < 0$ and $S\simeq g(\gamma = 1)$. For $L\ll l_{\rm eq}$, $S \propto L/l_{\rm eq}$. The coefficient functions $f(\gamma)$ and $g(\gamma)$ are given in the Supplemental Material ~\cite{SuppMat}.
}
\end{figure}

Substituting the results for $T_\pm(x)$ in Eq.~(\ref{eq:Noise2ModesMain}), we obtain three topologically distinct types of the scaling of noise $S$ in the incoherent regime, $L \gg l_{\rm eq}$, that were announced in the Introduction. These results are illustrated  in Fig.~\ref{fig:NoisePlots} where we choose a representative edge for each class. For the $\nu =3/5$ edge we have $\nu_Q = -1 < 0$ (antiballistic heat transport), so that the noise is constant for $L / l_{\rm eq} \gg 1$. For $\nu = 2/3$, the heat transport is diffusive in view of $\nu_Q = 0$, and the noise decays as $\sqrt{l_{\rm eq}/L}$. Finally, the boundary between the 4/3 and 1/5 fractions is characterized by $\nu_Q = 1$ and thus ballistic heat transport, so that the noise decays exponentially in $L / l_{\rm eq}$. In the short edge limit, $L\ll l_{\rm eq}$, the tunneling events are rare, and the noise scales as $S\propto L/l_{\rm eq}$ in all cases.

\textit{Discussion.--}
We expect our noise classification to be applicable not only for conventional edge structures, but also for experimentally engineered edge structures with equal $G$ but different $G^Q$~\cite{Grivnin2014,Ronen2018,Cohen2019,Lin2019}. For instance, an interface between $\nu=4/3$ and $\nu=2/3$ FQH states is expected to host two copropagating $1/3$ channels. In the incoherent regime, this edge yields $G\simeq (2/3)e^2/h$ and $G^Q\simeq 2\kappa T$ in contrast to the conventional $\nu=2/3$ edge where $G\simeq (2/3)e^2/h$ but $G^Q\simeq 0$~\cite{Protopopov2017,Nosiglia2018}. Our model predicts that only the latter edge generates nonvanishing noise, which can be tested experimentally.
Recent measurements of noise on conductance plateaux in QPC geometries~\cite{Bhattacharyya2019} could also be analyzed with our model. 

We further anticipate that our classification can be extended to the exotic non-Abelian $\nu=5/2$ state. Current research points towards three prominent and competing theories of the edge structure~\cite{Moore1991,Levin2007,Lee2007,Fidkowski2013,Son2015,Simon2018,Ma2019,Feldman2018} and our preliminary results~\cite{FutureNAb} indicate that these proposals fall into distinct noise classes (ballistic heat flow and exponentially suppressed noise for the Pfaffian and particle-hole-Pfaffian edges, and antiballistic heat flow and constant noise for the anti-Pfaffian edge;	 assuming no equilibration between the Landau levels). Our results thereby provide an experimental signature that may distinguish between these theories. 

Our analysis of the incoherent regime does not depend on microscopic parameters but only requires $L\gg l_{\rm eq}$. At the same time, the equilibration length $l_{\rm eq}$ depends on temperature, tunneling, and interaction between the channels (see Ref.~\cite{Protopopov2017} for the corresponding analysis for $\nu = 2/3$). 
For Abelian edges, experimental data imply a charge equilibration length $l_{\rm eq} \lesssim 10$ $\mu$m and a thermal length $\tilde{l}_{\rm eq} \sim 30$ $\mu$m at an ambient temperature $T_0 = 10$ mK~\cite{Banerjee2017,Banerjee2018,Aharon2019,Ma2019,Cohen2019}. This is consistent with our model where $l_{\rm eq}\sim \tilde{l}_{\rm eq}$. The data indicate further that edges with $L\approx 50$ $\mu$m are sufficiently short to suppress heat leakage into the bulk. Thus, measurements on edges with electrical and thermal equilibration and without thermal leakage should be feasible.

We end by noting that by measuring the constant noise (i.e. for $\nu_Q<0$) in the incoherent regime, the functions $f(\gamma)$ and $g(\gamma)$~\cite{SuppMat} (and their generalization to other edges in the same class) can be used to extract $\gamma$ and probe violations of the Wiedemann-Franz law. 

\textit{Summary.--}
We studied shot noise on incoherent fractional quantum Hall edges and found that the asymptotic characteristics of the shot noise can be classified into three topological universality classes. These classes represent the three different possibilities in the interplay between electrical and thermal transport: ballistic charge and heat transport generate either vanishing or exponentially suppressed noise, ballistic charge and diffusive heat transport generate algebraically decaying noise, and ballistic charge but antiballistic heat transport generate constant noise. We expect this classification to be applicable to conventional filling factors, edge states at the interface between different FQH bulk states, more complex QPC geometries, and non-Abelian states. Our findings suggest that noise measurements in the incoherent regime provide important insights into current edge state theories.

\begin{acknowledgments}
\textit{Acknowledgments.--}
We thank Moty Heiblum, Rajarshi Bhattacharyya, Amir Rosenblatt, and Roland Willa for useful discussions. C.S.,  Y.G., and  A.D.M. acknowledge support by DFG Grant No. MI 658/10-1. Y.G. further acknowledges support by  DFG RO 2247/11-1 and CRC 183 (project C01).  J.P. acknowledges support by the Koshland Foundation.   

\end{acknowledgments}


\begin{thebibliography}{53}
\bibitem [{gDe()}]{gDepNote}%
  \BibitemOpen
  \href@noop {} {\bibinfo  {journal} {For simplicity, the (Luttinger liquid
  like) temperature dependence of $g_{n,m}$ is ignored, since it does not lead
  to any qualitative changes in our results}\ }\BibitemShut {NoStop}%
\end{thebibliography}
%

\clearpage
\newpage

\onecolumngrid
\setcounter{equation}{0}
\setcounter{figure}{0}
\setcounter{table}{0}
\setcounter{page}{1}
\renewcommand{\theequation}{S\arabic{equation}}
\renewcommand{\thefigure}{S\arabic{figure}}
\renewcommand{\bibnumfmt}[1]{[S#1]}
\renewcommand{\citenumfont}[1]{S#1}

\bigskip

\begin{center}
\large{\bf Supplemental Material for "Topological Classification of Shot Noise on Fractional Quantum Hall Edges"\\}
\end{center}
\begin{center}
Christian Sp\r{a}nsl\"{a}tt,$^{1,2}$ Jinhong Park,$^{3}$ Yuval Gefen,$^{1,3}$ and Alexander D. Mirlin$^{1,2,4,5}$
\\
{\it $^1$Institut f\"{u}r Nanotechnologie, Karlsruhe Institute of Technology, 76021 Karlsruhe, Germany \\$^2$Institut f\"{u}r Theorie der Kondensierte Materie, Karlsruhe Institute of Technology, 76128 Karlsruhe, Germany \\
$^{3}$Department of Condensed Matter Physics, Weizmann Institute of Science, Rehovot 76100, Israel \\ $^{4}$Petersburg Nuclear Physics Institute, 188300 St. Petersburg, Russia
\\ $^{5}$L.\,D.~Landau Institute for Theoretical Physics RAS, 119334 Moscow, Russia}\\
(Dated: \today)

\end{center}

In this supplemental material, we provide details of the reduction of the general edge theory into an effective two-mode model with a single equilibration length. We also derive Eq.~\eqref{eq:Noise2ModesMain} for the noise in the main paper, together with the three types of the asymptotic behavior of the noise. We end by computing explicit expressions for the temperature profiles and the noise for a few representative edges. 

\section{\label{sec:Reduction} SA. Reduction to the effective theory}

To simplify the treatment of the general edge, we merge the edge channels into sets of the same chirality. The channels within these sets are assumed to equilibrate on the length scale $\sim a$ (which defines the shortest length scale of the problem) into  hydrodynamic modes. We start with the general matrix $\mathcal{M}_V$ in Eq.~\eqref{eq:VMatrix} in the main paper and require $V_n(x)=V_+(x)$ for $1\leq n\leq n_d$ (with chirality $\chi_+$) and $V_n(x)=V_-(x)$ for $n_d<n\leq N$ (chirality $\chi_-$). We obtain
\begin{equation}
\label{eq:VEffectiveEquation}
	\partial_{x}\begin{pmatrix}
V_{+}(x) \\[0.3cm] V_{-}(x) 
\end{pmatrix}=
\begin{pmatrix}
-\chi_+\sum_{n>n_d} \frac{g_{1,n}}{a \nu_1} & \chi_+\sum_{n>n_d} \frac{g_{1,n}}{a \nu_1}\\[0.3cm]
\chi_-\sum_{n\leq n_d} \frac{g_{n_d+1,n}}{a \chi_N \nu_N} & -\chi_-\sum_{n\leq n_d} \frac{g_{n_d+1,n}}{a \nu_N}
\end{pmatrix}
\begin{pmatrix}
V_{+}(x) \\[0.3cm] V_{-}(x) 
\end{pmatrix},
\end{equation}
together with the constraints
\begin{subequations}
\begin{align}
& \sum_{n>n_d} \frac{g_{1,n}}{\nu_1} = 	\sum_{n>n_d} \frac{g_{2,n}}{\nu_2} = \hdots = \sum_{n>n_d} \frac{g_{n_d,n}}{\nu_{n_d}}, \\
& \sum_{n\leq n_d} \frac{g_{n_d+1,n}}{\nu_{n_d+1}} = \sum_{n\leq n_d} \frac{g_{n_d+2,n}}{\nu_{n_d+2}} = \hdots = \sum_{n\leq n_d} \frac{g_{N,n}}{\nu_{N}}.
\end{align}
\end{subequations}
Inserting these constraints back into Eq.~\eqref{eq:VEffectiveEquation} results in an effective $2$-channel problem with
\begin{equation}
\label{eq:V2times2}
	\mathcal{M}_V = \frac{g_{+,-}}{a} \begin{pmatrix}
-\chi_+ \left( \sum_{n=1}^{n_d} \nu_n \right)^{-1} & \chi_+ \left( \sum_{n=1}^{n_d} \nu_n \right)^{-1}\\
\chi_- \left( \sum_{n=n_d+1}^{N} \nu_n \right)^{-1} & -\chi_- \left( \sum_{n=n_d+1}^{N} \nu_n \right)^{-1}
\end{pmatrix} \equiv \frac{1}{l_{+,-}} \begin{pmatrix}
-\chi_+/\nu_+ & \chi_+/\nu_+\\
\chi_-/\nu_- & -\chi_-/\nu_-
\end{pmatrix},
\end{equation}
where $\chi_{\pm}$ are the chiralities of the two hydrodynamical modes and $l_{+,-}\equiv a/g_{+,-}$ is a length related to equilibration between them. The exact dependence of $l_{+,-}$ on $g_{n,m}$ and the channel conductances is generally complicated. However, its exact expression is unimportant in the continuum limit and instead we use $l_{\rm eq} \equiv a\nu_+\nu_-/(g_{+,-}(\nu_+-\nu_-))$ (see Sec.~SB) as a phenomenological parameter. Physically, $l_{\rm eq}$ depends generally on the temperature and the microscopic structure of the edge~\cite{Protopopov2018SM}. Our results for the incoherent regime are however only dependent on the condition $L/l_{\rm eq}\gg 1$. 

We treat the temperature profiles in the same manner. The amount of heat carried by each Abelian channel is the same~\cite{Capelli2002SM} and the resulting effective matrix therefore reads 
\begin{equation}
\label{eq:T2times2}
	\mathcal{M}_{T\;} = \frac{\gamma}{l_{+,-}} \begin{pmatrix}
-\chi_+ n_d & \chi_+ n_d\\
\chi_- n_u & -\chi_- n_u
\end{pmatrix},
\end{equation}
where $l_{+,-}/\gamma$ parametrizes the effective thermal equilibration length. Here $\gamma$ is the Wiedemann-Franz parameter of the effective heat tunneling current~\cite{Nosiglia2018SM}. We also obtain $\Delta \vec{V}(x) = (V_+(x)-V_-(x))^2e^2/(h \kappa l_{+,-})(-1,1)^T$ (see Eq.~\eqref{eq:TemperatureEquation} in the main paper).

The merging of the channels is equivalent to assuming the following hierarchy of length scales
\begin{equation}
a \sim l_{-,-} \sim l_{+,+} \ll l_{+,-} \sim l_{\rm eq}  \ll L,
\label{eq:scale_hierarchy}
\end{equation}
where $l_{-,-}$, $l_{+,+}$, and $l_{+,-}$ are lengths corresponding to tunneling between modes within the $-$ set, within the $+$ set, and between the sets respectively. The only remaining length scale is therefore $l_{-,+} \sim l_{\rm eq}$ which we use to define the incoherent regime (see the last inequality in Eq.~\eqref{eq:scale_hierarchy}).

\section{SB. Voltage profiles and charge conductances for the effective counter-propagating modes}
\label{sec:SB}

\begin{figure}[t]
\includegraphics[width=0.75\textwidth]{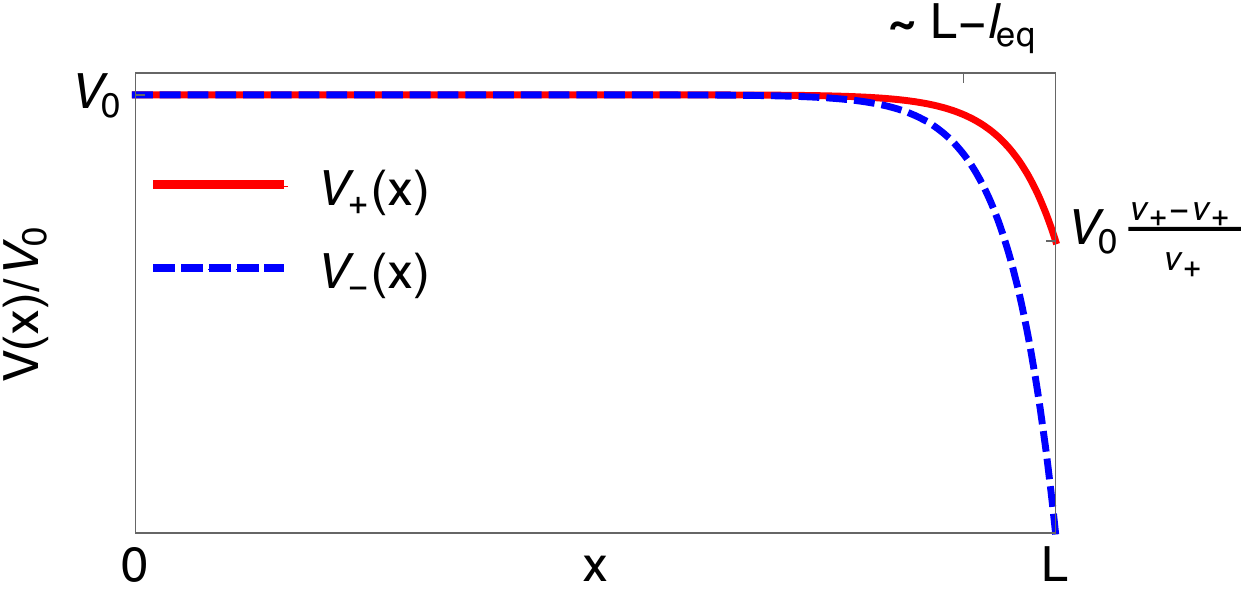}\caption{\label{fig:VoltageProfiles} (Color online) Voltage profiles for counter-propagating modes $V_+(x)$ (downstream, red line) and $V_-(x)$ (upstream, blue, dashed line) with filling factor discontinuities $\nu_+$ and $\nu_-<\nu_+$, respectively (see Fig.~\ref{fig:Model} in the main paper). The left contact, $C_{\rm down}$ ($x=0$), is biased with $V_0$ and the downstream mode voltage $V_+(x)$ drops from $V_0$ to $(\nu_+-\nu_-)V_0/\nu_+$ due to equilibration within the length $l_{\rm eq}$ from the right grounded contact, $C_{\rm up}$ ($x=L$). Here, we have used $L/l_{\rm eq}=20$.}
\end{figure}

We consider the voltage profiles for two counterpropagating hydrodynamic modes (the same results hold also for two simple edge channels) with filling factor discontinuities $\nu_+$ and $\nu_-<\nu_+$ and chiralities $\chi_\pm=\pm1$. Using the matrix in Eq.~\eqref{eq:V2times2} and boundary conditions $V_+(0)=V_0$ and $V_-(L)=0$, the solution of Eq.~\eqref{eq:VoltageEquation} in the main paper reads
\begin{subequations}
\label{eq:VProfiles}
	\begin{align}
		& V_+(x) = V_0\frac{\nu_+ e^{L/l_{\rm eq}}-\nu_- e^{x/l_{\rm eq}}}{\nu_+e^{L/l_{\rm eq}} - \nu_-}, \\
		& V_-(x) = V_0 \frac{\nu_+ e^{x/l_{\rm eq}}-\nu_+ e^{L/l_{\rm eq}}}{\nu_+e^{L/l_{\rm eq}} - \nu_-},
	\end{align}
\end{subequations}
where $l_{\rm eq} \equiv a\nu_+\nu_-/(g_{+,-}(\nu_+-\nu_-))$. These profiles are plotted in Fig.~\ref{fig:VoltageProfiles}. It follows that
\begin{equation}
\label{eq:Vdiff}
\Delta \vec{V}(x)=\frac{V_0^2 e^2}{h\kappa l_{+,-}}\frac{e^{2x/l_{\rm eq}}(\nu_+-\nu_-)^2}{(e^{L/l_{\rm eq}}\nu_+-\nu_-)^2} \begin{pmatrix}
	-1 \\
	+1
\end{pmatrix}.
\end{equation}
In the limit $L/l_{\rm eq}\rightarrow \infty$, we obtain $V_+(L)=V_0 (\nu_+-\nu_-)/\nu_+$ and the conductance $G_d$ determining the response of the current on this edge to the potential of the left contact $C_{\rm down}$ ($x=0$) reads $G_d\simeq (\nu_+-\nu_-)e^2/h$. Reversing the role of source and drain by choosing $V_+(0)=0$ and $V_-(L)=V_0$ yields for the conductance $G_u$ determining the response of the current on this edge to the potential of the right contact $C_{\rm up}$ ($x=L$)   in the same limit $G_u\simeq 0$. The two-terminal conductance $G$ that includes also the current flowing on the second edge connecting $C_{\rm down}$ and $C_{\rm up}$ is generically $G= G_u + G_d$. 
We thus have a quantized conductance in the incoherent regime 
\begin{equation}
G \simeq G_d\simeq (\nu_+-\nu_-)e^2/h,
\end{equation}
up to exponentially small corrections.
In contrast, for the perfectly clean edge, one finds $G_d = \nu_+ e^2/h$ and $G_u = \nu_- e^2/h$, yielding $G=(\nu_+ + \nu_-)e^2/h$. This result follows straight-forwardly from a Landau-B\"uttiker treatment in the absence of the inter-mode interaction but holds also for a clean interacting edge \cite{Protopopov2018SM}. 

\section{SC. Derivation of DC noise for the effective counter-propagating modes}

Within the approximation scheme that reduces the general edge structure to two hydrodynamical modes, Eq.~\eqref{eq:NoiseEquation} of the main paper reduces to
\begin{align}
\label{eq:NoiseEquation2Modes}
&\partial_{x}\begin{pmatrix}
\delta I_{+}(x) \\ \delta I_{-}(x) 
\end{pmatrix}
= \frac{1}{l_{+,-}} \begin{pmatrix}
-\chi_+/\nu_+ & \chi_-/\nu_-\\
\chi_+/\nu_+ & -\chi_-/\nu_-
\end{pmatrix}
\begin{pmatrix}
\delta I_{+}(x) \\ \delta I_{-}(x) 
\end{pmatrix}+\delta I_{+-}^{\tau,{\rm int}}(x)
\begin{pmatrix}
 -1\\ +1
\end{pmatrix} \notag \\[0.3cm]
& \Longleftrightarrow\ \ \partial_x\vec{\delta I}(x) = \mathcal{M}_I \vec{\delta I}(x) + \vec{\delta I}^{\tau,{\rm int}}(x).
\end{align}
This system of two coupled inhomogeneous linear differential equations has the general solution
\begin{equation}
\label{eq:deltaISol}
\vec{\delta I}(x) = e^{x \mathcal{M}_I} \vec{\delta I}(0) + \int_0^x dy\; e^{\mathcal{M}_I (x-y)} \vec{\delta I}^{\tau,{\rm int}}(y).
\end{equation}
To proceed, we first consider an edge where all channels have the same chirality (right-moving): $\chi_+=\chi_-=+1$. This is a maximally chiral edge in which we have split all modes in two sets. 
The appropriate boundary conditions read
$\delta I_{+}(0)=\delta I_{-}(0)=0$ assuming noiseless bias voltages in the contacts. The measured noise $S\equiv \overline{(\delta I_+(x)+\delta I_-(x))^2}$, then vanishes identically, $S=0$,  since $\delta I_+(x)=-\delta I_-(x)$ due to current conservation. Physically, this holds since the maximally chiral edge lacks any mechanism for current partitioning. We conclude that the maximally chiral edge generates identically zero noise.

Next, we return to an edge with counterpropagating modes. The appropriate boundary conditions are now $\delta I_{+}(0)=\delta I_{-}(L)=0$. The noise (which can be calculated in  $C_{\rm down}$, in $C_{\rm up}$, or at any other point along the edge),  $S\equiv \overline{ \delta I_+(L)^2}=\overline{ \delta I_-(0)^2}$,  becomes
\begin{equation}
\label{eq:Noise2Modes}
S \simeq \frac{2e^2}{h l_{\rm eq}}\frac{ \nu_-}{ \nu_+} (\nu_+-\nu_-) \int_0^L dx\;\frac{e^{-\frac{2x}{l_{\rm eq}}}k_B\left[T_+(x) +T_-(x) \right]}{(1-e^{-\frac{L}{l_{\rm eq}}}\nu_-/\nu_+)^2},
\end{equation}
which is Eq.~\eqref{eq:Noise2ModesMain} in the main paper. In this derivation, we have used the definition $l_{\rm eq} \equiv a\nu_+\nu_-/[g_{+,-}(\nu_+-\nu_-)]$ as well as the continuum limit of Eq.~\eqref{eq:TempFluct} in the main paper,
\begin{equation}
\overline{\delta I^{\tau,{\rm int}}_{+-}(x)\delta I^{\tau,{\rm int}}_{+-}(y)} = \frac{2e^2}{h}g_{+,-}k_B\left[T_{+}(x)+T_{-}(y)\right]\delta(x-y).
\end{equation}

From Eq.~\eqref{eq:Noise2Modes} we see that the noise depends crucially on the temperature profiles along the edge. The noise classification (see the main paper and the next section in this Supplemental Material) follows directly from the fact that there are three distinct types of temperature profiles corresponding to $\nu_Q = n_d-n_u$ being positive, zero, or negative. This property is fixed by the topological order of the FQH state.

We further find that $S$ is symmetric with respect to interchange of source and drain contacts: the result does not depend on whether $C_{\rm down}$ is biased and $C_{\rm up}$ is grounded or vice versa. In fact, the noise is only dependent on the absolute value of the difference of potentials of the two contacts. This is related to the fact 
that the positions of hot spot and of the noise spot are fixed by the net chirality of the edge and are thus independent on the values and signs of the applied potentials. 

\section{SD. Temperature profiles for the effective counter-propagating modes}

We first consider $\nu_Q>0 \Leftrightarrow n_d>n_u$. Using Eq.~\eqref{eq:T2times2} and \eqref{eq:Vdiff} in Eq.~\eqref{eq:TemperatureEquation} of the main paper, both temperature profile solutions take the form
\begin{equation}
\label{eq:TDecay}
k_B T_{\pm}(x) \sim V_0 \times \left[\mathcal{O}\left(e^{-\frac{\gamma}{2}\frac{\nu_+ \nu_-}{\nu_+-\nu_-}\frac{L-x}{l_{\rm eq}}}\right)+\mathcal{O}\left(e^{-\frac{L-x}{l_{\rm eq}}}\right)\right],
\end{equation}
up to boundary corrections. We have here used the boundary conditions $T_+(0)=T_-(L)=0$, corresponding to the assumption that the contact temperatures are zero. From~\eqref{eq:TDecay}, we see that the heating of the edge is proportional to the bias voltage $V_0$ and decays exponentially away from the hot spot $x\sim L$ towards $x=0$ on two competing length scales: $l_{\rm eq}$ and $2l_{\rm eq}(\nu_+-\nu_-)/(\gamma \nu_+ \nu_-)$ where the latter is typically dominating. The heat reaching the noise-generating spot is therefore exponentially suppressed in $L/l_{\rm eq}$. Using these temperature profiles in Eq.~\eqref{eq:Noise2Modes}, we obtain zero noise, up to exponentially small corrections in $L/l_{\rm eq}$:
\begin{equation}
	S\sim \exp\left(-\frac{\gamma}{2}\frac{\nu_+ \nu_-}{\nu_+-\nu_-}\frac{L}{l_{\rm eq}}\right).
\end{equation}
The special case with $n_u=0$ generates $T_{\pm}(x)=0$, i.e. the temperature along the edge equals that of the bias contact.

Next, we consider $\nu_Q=0 \Leftrightarrow n_u=n_d$ and find diffusive temperature profiles
\begin{equation}
\label{eq:TDiffusive}
k_B T_{\pm}(x) \sim V_0 \times \sqrt{x/L}.
\end{equation}
The heat reaching the noise-generating spot at $x \sim l_{\rm eq}$ decays slowly as $L/l_{\rm eq}$ increases. Such temperature profiles generate through Eq.~\eqref{eq:Noise2Modes} an asymptotic noise behaviour 
\begin{equation}
S\sim (e^3V_0/h) \times \sqrt{l_{\rm eq}/L}. 
\end{equation}
This result is consistent with the previous result~\cite{Park2019SM} for the $\nu=2/3$ edge. 

When $\nu_Q<0 \Leftrightarrow n_u>n_d$, we find
\begin{equation}
\label{eq:TConst}
k_B T_{\pm}(x) \sim V_0 \times {\rm const.},
\end{equation}
up to boundary effects. The noise becomes asymptotically constant as a function of $L/l_{\rm eq}$, 
\begin{equation}
S\sim  (e^3V_0/h) \times {\rm const}.
\end{equation}

The three types of temperature profiles can be anticipated from the matrix $\exp\left( x \mathcal{M}_T\right)$ entering Eq.~\eqref{eq:T2times2}. If $n_d=n_u$, the matrix elements are linear in $x$, otherwise exponential. This property together with appropriate boundary conditions determine the type of heat transport. 

Finally, it is instructive to verify that our formalism correctly reproduces the equilibrium noise. For this purpose, we consider a constant ambient temperature along the edge and in the contacts, $T_+(x)=T_-(x)=T$ for $0\leq x\leq L$, and no bias. Using the local fluctuation-dissipation relations $\overline{(\delta I_{\pm}(0))^2}=\nu_\pm 2 e ^2 k_B T/h$,  $\overline{(\delta I_{\pm}(L))^2}=\nu_{\pm}2e^2 k_B T/h$, and $\overline{\delta I^{\tau,{\rm int}}_{+,-}(x)\delta I^{\tau,{\rm int}}_{+-}(y)}=2e^2g_{+,-}k_B T\delta(x-y)/h$, we obtain in the limit $L/l_{\rm eq}\rightarrow \infty$
\begin{equation}
\label{eq:NJN}
S = \overline{(\delta I_{+}(L)+\delta I_{-}(L))^2} \simeq 2k_BT (\nu_+-\nu_-)\frac{e^2\nu_-}{h\nu_+} +2 k_B T(\nu_+-\nu_-)^2\frac{e^2}{h\nu_+} =  2Gk_BT,
\end{equation}
where $G=(\nu_+-\nu_-)e^2/h=\sum_n(\chi_n \nu_n)e^2/h=\nu e^2/h$ is the charge conductance of the equilibrated edge. Equation \eqref{eq:NJN} is nothing but the thermal Johnson-Nyquist noise on the edge. Adding the contribution from the other edge (the FQH geometry requires every contact to connect two edges), the total thermal noise becomes $4Gk_BT$, which is the proper value of the equilibrium noise as dictated by the fluctuation-dissipation theorem applied to the whole system. 

\section{SE. Explicit noise expressions}
\label{sec:noiseExpressions}

\begin{figure}[t]
\includegraphics[width=0.5\textwidth]{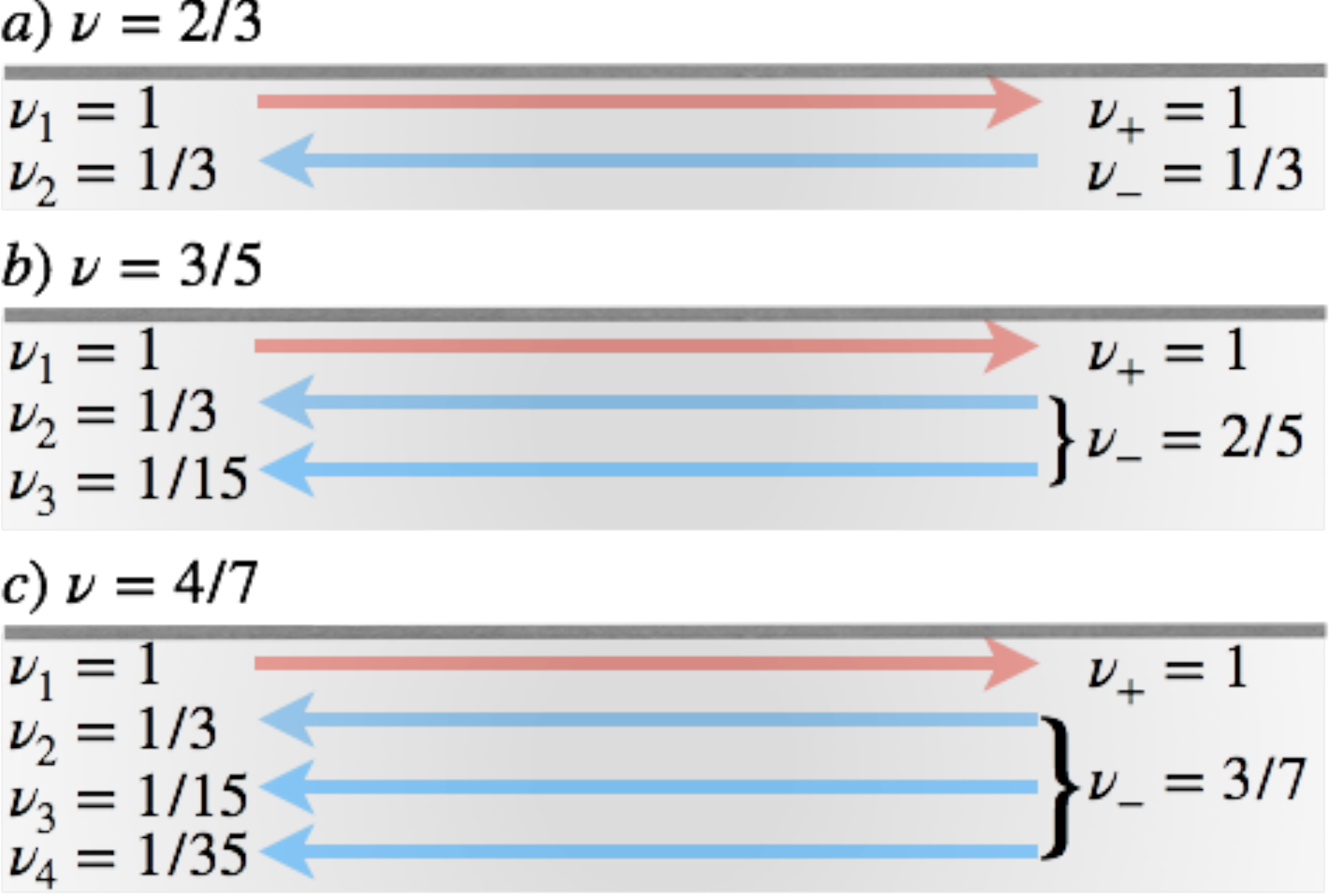}\caption{\label{fig:EdgeStructures} (Color online) FQH edge structures for a) $\nu=2/3$, b) $\nu=3/5$, and $\nu=4/7$. Downstream modes (chirality $\chi_+=+1$) are depicted in red and upstream (chirality $\chi_-=-1$) modes in blue. Modes with the same chirality are merged into effective hydrodynamic modes with effective filling factors $\nu_+$ and $\nu_-$ respectively.}
\end{figure}

In this section, we give explicit expressions for the the temperature profiles and the noise for a few representative examples of FQH edges with counterpropagating modes. 

\subsection{The $\nu=2/3$ edge}

For the $\nu=2/3$ edge, the mode structure of the edge corresponds  to $\nu_+=1$, $\nu_-=1/3$, $n_d=1$, and $n_u=1$ (see Fig.~\ref{fig:EdgeStructures}). This edge is  thus characterized by $\nu_Q =0$ and therefore  by diffusive heat transport. Solving Eq.~\eqref{eq:TemperatureEquation} of the main paper using Eq.~\eqref{eq:T2times2}, we find 
\begin{subequations}
	\begin{align}
		& k_B^2T_+^2(x) = \frac{3V_0^2 \left(\gamma  (\gamma +2) x e^{\frac{2 L}{l_{\rm eq}}}+(\gamma -2) \left(-e^{\frac{2 x}{l_{\rm eq}}} (2l_{\rm eq}+\gamma  L)+2l_{\rm eq}+\gamma  L-\gamma  x\right)\right)}{2 \pi^2  \left(1-3 e^{\frac{L}{l_{\rm eq}}}\right)^2 (2l_{\rm eq}+\gamma  L)}, \\
		& k_B^2T_-^2(x) = \frac{3V_0^2 \left(-(\gamma +2) e^{\frac{2x}{l_{\rm eq}}} (2l_{\rm eq}+\gamma  L)+(\gamma +2) e^{\frac{2L}{l_{\rm eq}}} (2l_{\rm eq}+\gamma  x)+(\gamma -2) \gamma  (L-x)\right)}{2 \pi^2 \left(1-3 e^{\frac{L}{l_{\rm eq}}}\right)^2 (2l_{\rm eq}+\gamma  L)}.
	\end{align}
\end{subequations}
The profiles $T^2_\pm(x)$ are linear in $x$ (up to boundary effects and exponentially small corrections), which indicates diffusive heat transport, in consistency with $\nu_Q=0$. Using these profiles in Eq.~\eqref{eq:Noise2Modes}, we obtain
\begin{equation}
	S \simeq f(\gamma)\frac{e^3 V_0}{h}\sqrt{\frac{l_{\rm eq}}{L}},
\end{equation}
where $f(\gamma)=\sqrt{2}\left[2\Gamma(3/2,4/\gamma)e^{4/\gamma} + \sqrt{\pi} \right]\sqrt{(2+\gamma)/6}/(18\pi)$ with $\Gamma(a,b)$ being the incomplete gamma function. This is (up to a factor $\sqrt{2}$ due to different definitions of $l_{\rm eq}$) the result obtained in Ref.~\onlinecite{Park2019SM}. The function $f(\gamma)$ is plotted in Fig.~\ref{fig:Noisefunctions}, and $f(\gamma=1)\approx 0.11$ is used in Fig.~\ref{fig:NoisePlots} in the main paper. It should be mentioned that $\gamma=9/5$ was derived for the non-interacting $\nu=2/3$ edge~\cite{Nosiglia2018SM}.

\subsection{The $\nu=3/5$ edge}

The $\nu=3/5$ edge is predicted to consist of three modes: one downstream $\nu_1=1$ mode and two upstream modes, $\nu_2=1/3$ and $\nu_3=1/15$. We therefore have $\nu_+=1$, $\nu_-=2/5$, $n_d=1$, and $n_u=2$ (see Fig.~\ref{fig:EdgeStructures}). This edge is  thus characterized by $\nu_Q =-1$, implying antiballistic heat transport. We find
\begin{subequations}
	\begin{align}
		k_B^2T_+^2(x) &= \frac{27 V_0^2 e^{-\frac{2\gamma  x}{3l_{\rm eq}}} }{(\gamma +3) \pi^2 \left(2-5 e^{\frac{L}{l_{\rm eq}}}\right)^2 \left(2 e^{\frac{2\gamma  L}{3l_{\rm eq}}}-1\right)}
		 \Bigg[ 2 (\gamma -1) e^{\frac{2\gamma  L}{3l_{\rm eq}}}-2 (\gamma +1)e^{\frac{2(\gamma +3) L}{3l_{\rm eq}}} -2 (\gamma -1) e^{\frac{2(\gamma  L+(\gamma +3)x)}{3l_{\rm eq}}} \notag \\
		&  +2 (\gamma +1) e^{\frac{2(x\gamma+(3+\gamma)L)}{3l_{\rm eq}}}-(\gamma -1) e^{\frac{2\gamma  x}{3l_{\rm eq}}}+(\gamma -1) e^{\frac{2(\gamma +3) x}{3l_{\rm eq}}}\Bigg], \\
		k_B^2T_-^2(x) &= \frac{27 V_0^2 e^{-\frac{2\gamma  x}{3l_{\rm eq}}} }{(\gamma +3) \pi^2 \left(2-5 e^{\frac{L}{l_{\rm eq}}}\right)^2 \left(2 e^{\frac{2\gamma  L}{3l_{\rm eq}}}-1\right)} 
		\Bigg[(\gamma -1) e^{\frac{2\gamma  L}{3l_{\rm eq}}}-(\gamma +1) e^{\frac{2(\gamma +3) L}{3l_{\rm eq}}}+2 (\gamma +1) e^{\frac{2(x\gamma+(\gamma +3)L)}{3l_{\rm eq}}} \notag \\
		& -2 (\gamma +1) e^{\frac{2(\gamma  L+(\gamma +3)x)}{3l_{\rm eq}}}-(\gamma -1) e^{\frac{2\gamma  x}{3l_{\rm eq}}}+(\gamma +1) e^{\frac{2(\gamma +3) x}{3l_{\rm eq}}}\Bigg].
	\end{align}
\end{subequations}

These profiles describe constant temperatures along the edge that drop sharply close to the contacts. The noise becomes
\begin{equation}
	S \simeq g(\gamma)\frac{e^3 V_0}{h},
\end{equation}
where $g(\gamma)=9\sqrt{\frac{(3+3\gamma)}{3+\gamma}}\Gamma(\frac{2+\gamma}{\gamma})\left(\sqrt{\pi} + 2 \Gamma(\frac{3}{2}+\frac{2}{\gamma}) _2\tilde{F}_1 (-\frac{1}{2},\frac{2}{\gamma},\frac{2+\gamma}{\gamma},\frac{1}{2}) \right)/(125 \pi \Gamma(\frac{3}{2}+\frac{2}{\gamma}))$. Here $\Gamma(a)$ is the gamma function and $_2 \tilde{F}_1(a,b,c,d)$ is the regularized hypergeometric function $_2 F_1(a,b,c,d)/\Gamma(c)$. We plot $g(\gamma)$ in Fig.~\ref{fig:Noisefunctions}, and $g(\gamma=1)\approx 0.07$ is used in Fig.~\ref{fig:NoisePlots} in the main paper.

\subsection{The $\nu=4/7$ edge} 

\begin{figure}[t]
\includegraphics[width=0.75\textwidth]{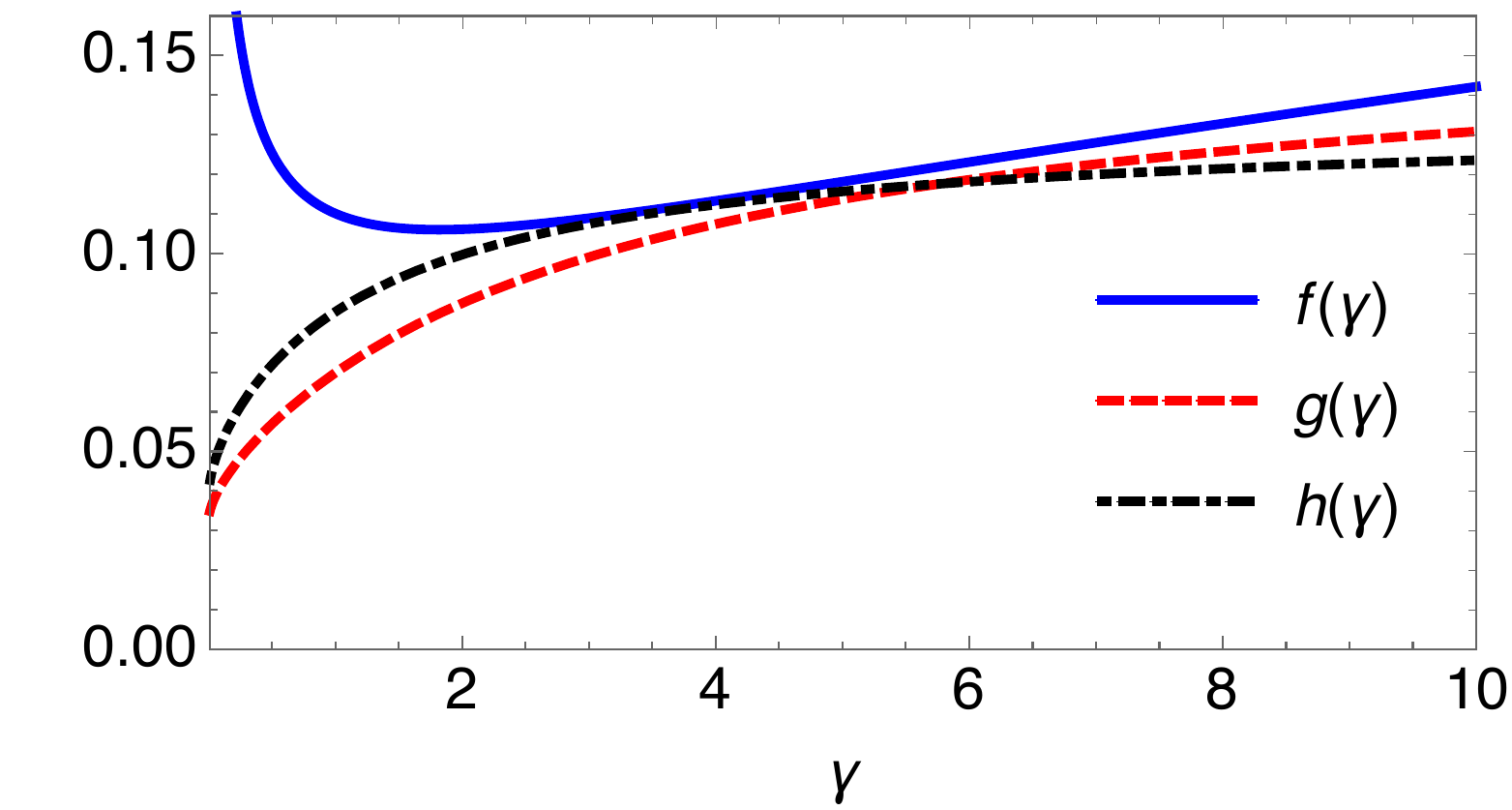}\caption{\label{fig:Noisefunctions} (Color online) The dimensionless functions $f(\gamma)$, $g(\gamma)$, and $h(\gamma)$ which determine the dependence of the noise (at some fixed $L/l_{\rm eq}\gg 1$) on the parameter $\gamma$ controlling the violation of the Wiedemann-Franz law in the inter-mode tunneling. For the exact expressions $f(\gamma)$, $g(\gamma)$, and $h(\gamma)$, see Sec. SE.}

\end{figure}

The $\nu=4/7$ edge is predicted to consist of four modes: one downstream $\nu_1=1$ mode and upstream modes $\nu_2=1/3$, $\nu_3=1/15$, and $\nu_4=1/35$ (see Fig.~\ref{fig:EdgeStructures}). We have therefore $\nu_+=1$, $\nu_-=3/7$, $n_d=1$, and $n_u=3$. This edge is thus characterized by $\nu_Q =-2$ and therefore by  antiballistic heat transport.  We obtain 
\begin{subequations}
	\begin{align}
		k_B^2T_+^2(x) &= \frac{36 V_0^2 e^{-\frac{3 \gamma  x}{2l_{\rm eq}}}}{(3 \gamma +4)\pi^2  \left(3-7 e^{\frac{L}{l_{\rm eq}}}\right)^2 \left(3 e^{\frac{3\gamma  L}{2l_{\rm eq}}}-1\right)}
	\Bigg[-3 (3 \gamma +2) e^{\frac{(3 \gamma +4) L}{2 l_{\rm eq}}}+(9 \gamma -6) e^{\frac{3 \gamma  L}{2 l_{\rm eq}}}\notag \\
& +3 (3 \gamma +2) e^{\frac{3 \gamma  (L+x)+4 L}{2 l_{\rm eq}}}+(6-9 \gamma ) e^{\frac{3 \gamma  (L+x)+4 x}{2 l_{\rm eq}}}+(3 \gamma -2) e^{\frac{(3 \gamma +4) x}{2 l_{\rm eq}}}+(2-3 \gamma ) e^{\frac{3 \gamma  x}{2 l_{\rm eq}}} \Bigg], \\
		k_B^2T_-^2(x) &= \frac{36 V_0^2 e^{-\frac{3 \gamma  x}{2l_{\rm eq}}}}{(3 \gamma +4)\pi^2  \left(3-7 e^{\frac{L}{l_{\rm eq}}}\right)^2 \left(3 e^{\frac{3\gamma  L}{2l_{\rm eq}}}-1\right)} 
		\Bigg[(3 \gamma -2) e^{\frac{3 \gamma  L}{2 l_{\rm eq}}}-(3 \gamma +2) e^{\frac{(3 \gamma +4) L}{2 l_{\rm eq}}}\notag \\
& +3 (3 \gamma +2) e^{\frac{3 \gamma  (L+x)+4 L}{2 l_{\rm eq}}}-3 (3 \gamma +2) e^{\frac{3 \gamma  (L+x)+4 x}{2 l_{\rm eq}}}+(3 \gamma +2) e^{\frac{(3 \gamma +4) x}{2 l_{\rm eq}}}+(2-3 \gamma ) e^{\frac{3 \gamma  x}{2 l_{\rm eq}}} \Bigg].
	\end{align}
\end{subequations}
Similarly to the $\nu=3/5$ edge, the temperature profiles are constant up to boundary corrections. The noise reads
\begin{equation}
	S \simeq h(\gamma)\frac{e^3 V_0}{h},
\end{equation}
with $h(\gamma)=36\sqrt{\frac{(2+3\gamma)}{4+3\gamma}}\Gamma(\frac{1+\gamma}{\gamma})\left(\sqrt{\pi} + 2 \Gamma(\frac{3}{2}+\frac{1}{\gamma}) _2\tilde{F}_1 (-\frac{1}{2},\frac{1}{\gamma},\frac{1+\gamma}{\gamma},\frac{1}{3}) \right)/(343 \pi \Gamma(\frac{3}{2}+\frac{1}{\gamma}))$. We plot $h(\gamma)$ in Fig.~\ref{fig:Noisefunctions}. 
\newpage
\clearpage

\end{document}